\documentstyle[sprocl,epsfig]{article}

\catcode`@=11
\def\citer{\@ifnextchar [{\@tempswatrue\@citexr}{\@tempswafalse\@citexr[]}}
 
\def\@citexr[#1]#2{\if@filesw\immediate\write\@auxout{\string\citation{#2}}\fi
  \def\@citea{}\@cite{\@for\@citeb:=#2\do
    {\@citea\def\@citea{--\penalty\@m}\@ifundefined
       {b@\@citeb}{{\bf ?}\@warning
       {Citation `\@citeb' on page \thepage \space undefined}}%
\hbox{\csname b@\@citeb\endcsname}}}{#1}}
\catcode`@=12

\def\reffi#1{\mbox{Fig.~\ref{#1}}}
\def\citere#1{\mbox{Ref.~\cite{#1}}}

\newcommand{\msbar}{$\overline{\rm{MS}}$}

\newcommand{\cp}{{\cal CP}}

\newcommand{\MW}{M_W}

\newcommand{\MA}{M_A}
\newcommand{\mh}{m_h}

\newcommand{\mhmax}{m_h^{\rm max}}

\newcommand{\mt}{m_t}
\newcommand{\mtexp}{m_t}

\newcommand{\tsf}{\theta\kern-.20em_{\tilde{f}}}
\newcommand{\tsfp}{\theta\kern-.20em_{\tilde{f}\prime}}
\newcommand{\tsq}{\theta\kern-.15em_{\tilde{q}}}

\newcommand{\sweff}{\sin^2\theta_{\mathrm{eff}}}

\newcommand{\VL}{\left( \begin{array}{c}}
\newcommand{\VR}{\end{array} \right)}
\newcommand{\ML}{\left( \begin{array}{cc}}
\newcommand{\MLd}{\left( \begin{array}{ccc}}
\newcommand{\MLv}{\left( \begin{array}{cccc}}
\newcommand{\MR}{\end{array} \right)}

\newcommand{\tb}{\tan \beta}

\newcommand{\gev}{\,\, \mathrm{GeV}}
\newcommand{\mev}{\,\, \mathrm{MeV}}

\newcommand{\BC}{\begin{center}}
\newcommand{\EC}{\end{center}}
\newcommand{\BE}{\begin{equation}}
\newcommand{\EE}{\end{equation}}
\newcommand{\BEA}{\begin{eqnarray}}
\newcommand{\BEAnn}{\begin{eqnarray*}}
\newcommand{\EEA}{\end{eqnarray}}
\newcommand{\EEAnn}{\end{eqnarray*}}

\newcommand{\id}{{\rm 1\kern-.12em
\rule{0.3pt}{1.5ex}\raisebox{0.0ex}{\rule{0.1em}{0.3pt}}}}
\newcommand{\lsim}
{\;\raisebox{-.3em}{$\stackrel{\displaystyle <}{\sim}$}\;}

\def\de{\delta}

\def\De{\Delta}

\begin{document}

\title{IMPACT OF A PRECISE TOP MASS MEASUREMENT
\footnote{Talk given by G.W.\ at LCWS04, Paris, April 2004.}
}

\author{S.~HEINEMEYER$^1$, S.~KRAML$^{1,2}$, W.~POROD$^3$,
G.~WEIGLEIN$^4$}

\address{\mbox{}\\[-.5em]
$^1$ CERN, TH Division, Dept.\ of Physics, 
CH-1211 Geneva 23, Switzerland\\[.3em]
$^b$ Institut f\"ur Hochenergiephysik, \"Osterr. Akademie d.
Wissenschaften,\\
A-1050 Vienna, Austria\\[.3em]
$^c$ Institut f\"ur Theoretische Physik, Universit\"at Z\"urich,\\
CH-8057 Z\"urich, Switzerland\\[.3em]
$^d$ IPPP, University of Durham, Durham DH1~3LE, UK
}


\maketitle\abstracts{
The physics impact of a precise determination of the top-quark mass,
$\mt$, at the Linear Collider (LC) is
discussed, and the results are compared with the prospective accuracy at
the LHC. The importance of a precise knowledge of $\mt$ for electroweak
precision observables and for Higgs physics in the MSSM is pointed out
in particular. We find that going from hadron collider to LC accuracy in
$\mt$ leads to an improvement of the investigated
quantities by up to an order of magnitude.
}
  
%***********************************************************************

\section{Introduction}

The mass of the top quark, $\mt$, is a fundamental parameter of the 
electroweak theory. It is by far the heaviest of all quark masses and
it is also larger than the masses of all other known fundamental particles. 
The large value of $\mt$ gives rise to a large coupling between the top 
quark and the Higgs boson and is furthermore important for flavour
physics. It could therefore provide a window to new physics. The correct
prediction of $\mt$ will be a crucial test for any fundamental theory.
The top-quark mass also plays an important role in electroweak precision
physics, as a consequence in particular of non-decoupling effects being
proportional to powers of $\mt$. A precise knowledge of $\mt$ is
therefore indispensable in order to have sensitivity to possible effects
of new physics in electroweak precision tests.

The current world average for the top-quark mass is $\mt = 178.0 \pm
4.3$~GeV~\cite{Azzi:2004rc,natured0}. The prospective accuracy at the 
Tevatron and the LHC is $\de\mtexp = \mbox{1--2}$~GeV~\cite{mtdetLHC}, 
while at the LC a very precise determination of $\mt$ with an accuracy
of $\de\mt \lsim 100 \mev$ will be 
possible~\cite{lctdrs,mtdet}.
This error contains both the experimental error of the mass parameter
extracted from the $t \bar t$ threshold measurements at the LC and 
the envisaged theoretical uncertainty from its transition into a suitable
short-distance mass (like the \msbar\ mass).

In the following some examples of the impact of a precise determination
of $\mt$ are discussed. More details can be found in \citere{deltamt}.

\section{Electroweak Precision Observables}

Electroweak precision observables (EWPO) can be used to
perform internal consistency checks of the model under consideration and
to obtain indirect constraints on unknown model parameters.
This is done by comparing experimental results of the EWPO with
their theory prediction within, for example, the Standard Model (SM) or
its minimal supersymmetric extension (MSSM).

There are two sources of theoretical uncertainties: those from
unknown higher-order corrections (``intrinsic'' theoretical
uncertainties), and those from experimental errors of the input
parameters (``parametric'' theoretical uncertainties). The
intrinsic uncertainties within the SM are
%currently~\cite{mwsweff}
%\BE
$\De\MW^{\rm intr,today} \approx 4 \mev$, %\quad
$\De\sweff^{\rm intr,today} \approx 5 \times 10^{-5}$
at present~\cite{mwsweff}.
%\label{eq:intruncSM}
%\EE
The parametric uncertainties induced by the current experimental error
of $\mt$ are $\De\MW^{\rm para,today} \approx 26 \mev$ and
$\De\sweff^{\rm para,today} \approx 14 \times 10^{-5}$. They are 
larger than the uncertainties induced by the experimental errors of all
other input parameters and are almost as large as the current
experimental errors of $\MW$ and $\sweff$. A future experimental error
of $\de\mt \approx 1.5$~GeV at the LHC will give rise to parametric
uncertainties of $\De\MW^{\rm para,LHC} \approx 9 \mev$,
$\De\sweff^{\rm para,LHC} \approx 4.5 \times 10^{-5}$, while the 
LC precision of $\de\mt \approx 0.1$~GeV will reduce the parametric
uncertainties to $\De\MW^{\rm para,LC} \approx 1 \mev$,
$\De\sweff^{\rm para,LC} \approx 0.3 \times 10^{-5}$. A comparison
with the parametric uncertainties induced by 
the other input parameters~\cite{deltamt} shows that the LC accuracy on
$\mt$ will be necessary in order to keep the parametric error induced by
$\mt$ at or below the level of the other uncertainties.
With the LHC accuracy on $\mt$, on the other hand, $\de\mt$ will be the
dominant source of uncertainty.

     \begin{figure}[ht] 
     \begin{center}
     \vspace*{.2cm}
     \epsfig{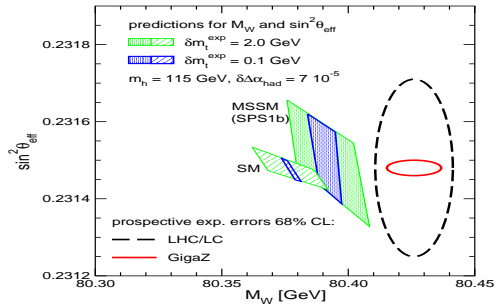}
     \end{center}
\vspace{-0.5em}
\caption{Predictions for $\MW$ and $\sweff$ in the SM and the MSSM
(SPS~1b). The inner
(blue) areas correspond to $\de\mtexp = 0.1 \gev$ (LC), while
the outer (green) areas arise from $\de\mtexp = 2 \gev$ (LHC).
The anticipated experimental errors on $\MW$ and $\sweff$ at the
LHC/LC and at a LC with GigaZ option are indicated.}
     \label{fig:mwsw}
\vspace{-0.5em}
     \end{figure}

In \reffi{fig:mwsw} the predictions for $\MW$ and $\sweff$ in the SM and
the MSSM are shown in comparison with the prospective experimental
accuracy obtainable at the LHC and a LC with
GigaZ option (low-energy running at the $Z$-boson
resonance and the $WW$-threshold). The MSSM parameters have
been chosen in this example according to the
reference point SPS~1b~\cite{sps}, and all SUSY parameters have been
varied within realistic error intervals. 
The figure shows that the improvement in $\de\mt$ from
$\de\mt = 2 \gev$ to $\de\mt = 0.1 \gev$ strongly reduces the
parametric uncertainty in the prediction for the EWPO.
In the SM case it leads to a reduction by about a factor of 10
in the allowed parameter space of the $\MW$--$\sweff$ plane.
In the MSSM case, where many additional parametric uncertainties enter,
a reduction by a factor of more than 2 is obtained in this example.
This precision will be crucial to
establish effects of new physics via EWPO.

\section{Implications For The MSSM}

In contrast to the SM, where the Higgs-boson mass is a free input
parameter,
the mass of the lightest $\cp$-even Higgs boson in the MSSM, $\mh$, can be
predicted in terms of other parameters of the model. While the
tree-level prediction for $\mh$ arises from the gauge sector of the
theory, large Yukawa corrections from the top and scalar top sector 
(for large values of $\tb$, the ratio of the vacuum expectation values
of the two Higgs doublets, also from the bottom and scalar bottom
sector) enter at the loop level. The leading one-loop correction
is proportional to $\mt^4$. The one-loop corrections can shift $\mh$ by
50--100\%. 

Since these very large corrections are proportional to
the fourth power of the top-quark mass, the predictions for $\mh$ and
many other observables in the MSSM Higgs sector strongly depend
on the precise value of $\mt$. As a rule of thumb~\cite{tbexcl}, a shift of
$\de \mt = 1\gev$ induces a parametric theoretical uncertainty of $\mh$
of also about $1 \gev$, i.e.\ $\De\mh^{\de\mt} \approx \de\mtexp$.

\begin{figure}[ht]
     \begin{center}
     \begin{tabular}{cc}
     \mbox{\epsfig{file=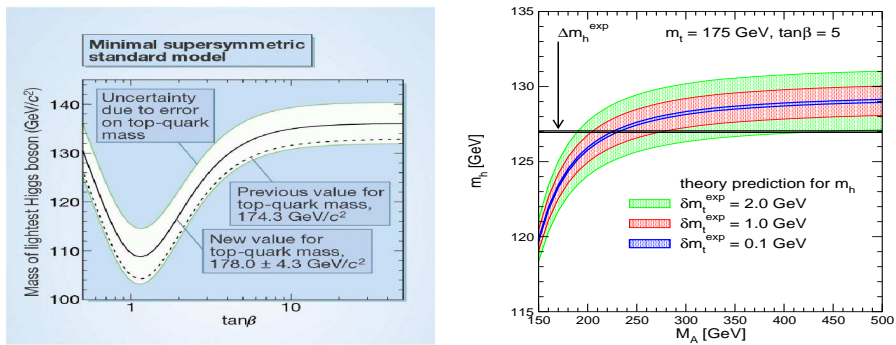,width=5.7cm,height=4.5cm}}&
     \mbox{\epsfig{file=mhMA06.cl.eps,width=5.7cm,height=4.5cm}}
     \end{tabular}
     \end{center}
\vspace{-0.5em}
\caption{Prediction for $\mh$ in the $\mhmax$ scenario of the MSSM as a
function of $\tb$ (left) and the mass of the $\cp$-odd Higgs boson,
$\MA$ (right). In the left plot %~\cite{naturetop}
the impact of the present experimental error of $\mt$ on the $\mh$
prediction is shown. The three bands in the right plot %~\cite{deltamt}
correspond to $\de\mtexp = 1, 2 \gev$ (LHC) and $\de\mtexp = 0.1 \gev$ (LC).
The anticipated experimental error on $\mh$ at the LC
is also indicated.
}
     \label{fig:mhmssm}
\vspace{-0.5em}
     \end{figure}

In \reffi{fig:mhmssm} the impact of the experimental error of $\mt$ on
the prediction for $\mh$ in the MSSM is shown. The parameters are chosen
according to the $\mhmax$ benchmark scenario~\cite{benchmark}. The band
in the left plot~\cite{naturetop} corresponds to the present
experimental error of $\mt$~\cite{Azzi:2004rc,natured0}, while in the 
right plot the situation at the LHC ($\de\mtexp = 1, 2 \gev$) is compared 
to the LC
($\de\mtexp = 0.1 \gev$). The figure shows that the LC precision on
$\mt$ will be necessary in order to match the experimental precision of
the $\mh$ determination with the accuracy of the theory prediction
(assuming that the intrinsic theoretical uncertainty can be reduced to
the same level, see \citere{mhiggsAEC}).

Further examples of the importance of 
a precise determination of $\mt$ in the MSSM are
the prediction of sparticle masses, parameter
determinations, and the reconstruction of the supersymmetric high scale
theory~\cite{deltamt}.

\end{document}